\documentclass[apj,numberedappendix]{emulateapj}
\usepackage{apjfonts}
\shorttitle{ENTROPY FROM QUASAR ACTIVITY} \shortauthors{LAPI,
CAVALIERE \& MENCI}
\journalinfo{Accepted by The Astrophysical Journal}
\begin{document}
\title{Intracluster and Intragroup Entropy from Quasar Activity}
%
\author{A. Lapi and A. Cavaliere}
\affil{Dip. Fisica, Univ. ``Tor Vergata'', Via Ricerca Scientifica
1, I-00133 Roma, Italy} \and
\author{N. Menci}
\affil{INAF, Oss. Astr. di Roma, Via Frascati 33, I-00044
Monteporzio Catone, Italy}
%
\begin{abstract}
We investigate how the hierarchical merging of dark matter halos,
the radiative cooling of baryons, and the energy feedback from
supernovae and active galactic nuclei or quasars combine to govern
the amount and the thermal state of the hot plasma pervading
groups and clusters of galaxies. We show that, by itself,
supernova preheating of the external gas flowing into clusters
falls short of explaining the observed X-ray scaling relations of
the plasma luminosity $L_X$ or the plasma entropy $K$ versus the
X-ray temperature $T$. To account for the scaling laws from rich
to poor clusters takes preheating enhanced by the energy input
from active galactic nuclei. In groups, on the other hand, the
internal impacts of powerful quasars going off in member galaxies
can blow some plasma out of the structure. So they depress $L_X$
and raise $K$ to the observed average levels; meanwhile, the
sporadic nature of such impulsive events generates the intrinsic
component of the wide scatter apparent in the data. The same
quasar feedback gives rise in groups to entropy profiles as steep
as observed, a feature hard to explain with simple preheating
schemes. Finally, we argue a close connection of the $L_X-T$ or
the $K-T$ relation with the $M_{\bullet}-\sigma$ correlation
between the host velocity dispersion and the masses of the black
holes, relics of the quasar activity.
\end{abstract}
%
\section{Introduction}
%
The hot medium pervading many single galaxies and most groups and
clusters shines in X rays by thermal bremsstrahlung and line
emission, see Sarazin (1988). Simple conditions are found to
prevail in rich clusters.

These emit huge powers $L_X\propto n^2\,\sqrt{T}\, R^3\sim
10^{44}-10^{45}$ erg s$^{-1}$ in X rays; the temperatures
$kT\approx 5$ keV, measured from the continuum and from high
excitation lines, are close to the virial values $kT_v\approx G\,
M\, m_p/10\, R$ in the gravitational wells mainly provided by dark
matter (DM) masses $M \approx 10^{15}\, M_{\odot}$ within sizes
$R$ of a few Mpcs \footnote{Throughout the paper $k$ is the
Boltzmann constant, $G$ is the gravitational constant, $m_p$ is
the proton mass, and $e$ is the electron charge.}. The inferred
gas number densities decline outward from central values $n\approx
10^{-3}$ cm$^{-3}$; so this medium with low $n$ and high $T$
satisfying $kT/ e^{2}\, n^{1/3}\sim 10^{12}$ constitutes the best
ion-electron plasma in the Universe ever, the intracluster
\emph{plasma} or ICP.

Such a medium is apparently {\it simple} on the following
accounts. Microscopically, it is constituted by pointlike
particles in thermal equilibrium. At the macroscopic end, the
overall baryonic fractions resulting from the ICP densities and
radial distributions inventoried in many clusters (White et al.
1993, Allen \& Fabian 1994) come to values $m/M \approx 0.16$;
this is close to the cosmic ratio $\Omega_b/\Omega_M$ of baryons
to DM obtained in the current Concordance Cosmology from the
parameters $\Omega_b\approx 0.044$ and $\Omega_M\approx 0.23$ (see
Bennett et al. 2003) \footnote{This will be used throughout the
paper, with its additional parameters $\Omega_{\Lambda}\approx
0.73$ for the dark energy density, and $H_0\approx 70\,
\mathrm{h}_{70}$ km s$^{-1}$ Mpc$^{-1}$ for the Hubble constant.}.
In addition, the chemical composition is reasonably constant from
cluster to cluster, and close to one-third of the solar value (see
Matteucci 2003).

However, surprises arise in moving from rich clusters toward poor
groups. In fact, similarly simple conditions holding in the
intragroup plasma (IGP) would imply that the luminosities retain
the gravitational scaling $L_X\propto T^2$ (Kaiser 1986). This
would apply if the IGP passively followed the DM evolution, and
retained the key cluster behaviors: $m/M\approx$ const, i.e.,
densities $n$ proportional to the DM mass density $\rho$; and
temperatures $T$ close to the virial value $T_v\propto M/R\propto
M^{2/3}\, \rho^{1/3}$.

Instead, the luminosities recently detected or revised (Horner
2001; O' Sullivan, Ponman \& Collins 2003; Osmond \& Ponman 2004)
are lower by factors of $10^{-1}$ to $10^{-2}$, see Fig.~1. The
figure also shows how the emissions from poor groups and large
galaxies scatter widely and often downward, a feature of a largely
intrinsic nature (Mushotzky 2004).

So in such smaller structures the plasma is surprisingly
underluminous and hence \emph{underdense}. This is an even more
surprising result, considering that in the standard hierarchical
cosmogony (see Peebles 1993) such earlier condensations ought to
be denser, if anything. Moreover, for $kT < 2$ keV the pinch of
highly excited metals contributes important line emissions that
imply a flatter $L_X\propto T$, if anything. How the observed
steep decrease may come about constitutes a widely debated issue.

Our proposal centers on the energy gained or lost by the baryons
through several processes: the gravitational heating driven by the
merging events that punctuate the hierarchical growth of DM
condensations (``halos'', see Peebles 1993); the radiative cooling
of the baryons; and the energy fed back to baryons when they
partly condense within galaxies into massive stars then exploding
as Type II supernovae (SNe), or accrete onto a central
supermassive black hole (BH) energizing an active galactic nucleus
(AGN) or a quasar.

The paper is organized as follows. In \S~2 we use a telling
quantity, the plasma entropy, to show that energy feedback from
astrophysical sources is needed to explain its high levels in poor
clusters and groups. In \S~3 we show that preheating by SNe alone
is not enough. In \S~4 we consider preheating enhanced by AGNs,
compute the resulting X-ray scaling relations, and critically
discuss the outcomes of this approach. In \S~5 we also consider
the internal impacts from quasars; we compute their effect on the
X-ray scaling relations and on the entropy profiles in groups, and
show how these solve the shortcomings of all external preheating
scenarios. In \S~6 we highlight and discuss the main features of
our comprehensive approach.

Auxiliary computations and derivations are given in the
Appendices. In Appendix A we derive Eq.~(2) of the main text from
the hydrostatic equilibrium, and give handy approximations. In
Appendix B we reformulate the classic Rankine-Hugoniot jump
conditions in a general form that is also valid for accretion
shocks/layers, and derive Eqs.~(5) and (6) of the main text. In
Appendix C we develop a new family of self-similar hydrodynamic
solutions describing the blast waves driven by the internal
impacts of quasars, and extensively used in \S~5.

%
\section{X-ray luminosity and entropy}
%

This paper deals with the energy budget of the baryons. The latter
experience \emph{gravitational} heating to $T\approx T_v$
(discussed in detail in \S~4) as they fall into the hierarchically
growing DM structures. \emph{Nongravitational} energy losses or
inputs deplete the baryon density; this is because losses trigger
baryon condensation to stars, while inputs cause outflow from and
hinder inflow into newly forming structures.

All such processes are probed with the \emph{adiabat} $K\equiv kT\,
n^{-2/3}\propto e^{2\, s/3\, k}$, a direct measure of the specific
entropy $s$ (see Bower 1997; Balogh, Babul \& Patton 1999). The
levels of the adiabat $K$ are linked to $L_X$ by the inverse
relation
\begin{equation}
K\propto L_X^{-1/3}\, T^{5/3}~,
\end{equation}
which obtains at $z\approx 0$ on eliminating $n$ between their
respective expressions; note that $T^{5/3}$ goes over to $T^{4/3}$
for important line emission, and that we are neglecting the weakly
$T$-dependent shape factors for $K$ and $L_X^{1/3}$.

Clearly, $K$ stays constant under adiabatic transformations of the
plasma. Gravitational heating would set the scaling $K\propto T$
(corresponding to $L_X\propto T^2$ after Eq.~[1]), but Fig.~2
shows the data for decreasing $T$ to deviate substantially
upwards; this indicates that additional nongravitational processes
occurred during a structure's merging history. The present paper
investigates how these processes affect the adiabat $K = K_2\,
\kappa(r)$, namely, the level $K_2$ at the virial radius $r=R$,
and the inner profile $\kappa(r)$.

\begin{figure}
\epsscale{1.0}\plotone{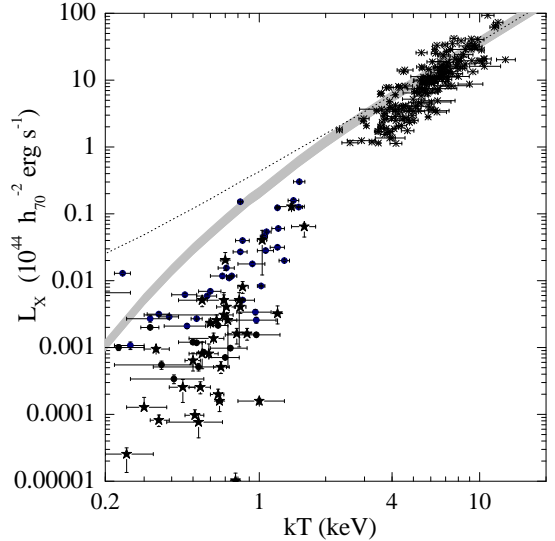}\caption{Integrated X-ray
luminosity $L_X$ vs. X-ray temperature $T$. Data for clusters
(\textit{crosses}) are from Horner (2001), for groups
(\textit{circles}) from Osmond \& Ponman (2004), and for
early-type galaxies (\textit{stars}) from O'Sullivan et al.
(2003). The \textit{dotted line} represents the gravitational
scaling, with line-emission included. The \textit{strip} (with 2
$\sigma$ width provided by the merging histories) illustrates our
results for SN preheating with $k\Delta T = \case{1}{4}$ keV per
particle, as discussed in \S~3.}
\end{figure}

The relation of these quantities to the density run $n(r)$ of the
plasma in hydrostatic equilibrium is derived in Appendix A, and
reads
\begin{equation}
{n(r)\over n_2} = {\kappa^{-3/5}(r)}\, \left[1+{2\over 5}\, \beta
\int_r^R{dr'}\, {d\phi\over dr'}\, \kappa^{-3/5}(r')\right]^{3/2}~;
\end{equation}
the boundary condition at $r=R$ is given by $n_2 = (kT_2/
K_2)^{3/2}$, but $K_2$ will be related to $T_2$ in \S~4. The
parameter $\beta=T_v/T_2$ is the ratio of the DM to the thermal
plasma scale height in the gravitational potential $\phi(r)$; the
latter is normalized to the one-dimensional dispersion $\sigma^2
\equiv kT_v/\mu m_p$, with $\mu \approx 0.6$ for the nearly cosmic
composition of the plasma (Cavaliere \& Fusco-Femiano 1976). For a
polytropic entropy distribution $\kappa(r)\propto
n(r)^{\Gamma-5/3}$ with uniform index $\Gamma\equiv 5/3 + d\ln
\kappa/d\ln n$, Eq.~(2) yields the familiar form $n = n_2\,
[1+(\Gamma-1)\, \beta\, \Delta \phi/\Gamma]^{1/(\Gamma-1)}$ in
terms of the potential drop $\Delta \phi$ inward of $R$; the
isothermal limit $n=n_2\, e^{\beta\, \Delta\phi}$ for $\Gamma=1$
provides the standard model to fit the X-ray surface brightness
profiles, which yields values $\beta\approx 0.7$ in rich clusters.
In particular, we use Eq.~(2) to compute integrated luminosities
$L_X\propto \int{dr}\, r^2 n^2(r)\, T^{1/2}(r)$ and central
entropies $K_{0.1}$ at $r=0.1\, R$, that we compare with the data.

Throughout the paper we will make use of ``semianalytic''
techniques. This is because the nongravitational processes affecting
$K$ include energy inputs and radiative losses, which interplay in
complex patterns with gravitational heating; so even numerical
simulations based on advanced $N$-body and hydrocodes are driven to,
or beyond their present limits, and have to borrow from semianalytic
models much subgrid physics (see discussion by Borgani et al. 2002).

Concerning radiative losses, they do operate within galaxies to
remove low-entropy gas by condensing it into stars, a process that
we include in our semianalytic modeling. But extensive cooling as
needed to substantially raise the residual ICP/IGP entropy or
depress $L_X$ would produce too many, unseen stars (see Voit \&
Bryan 2001; Sanderson \& Ponman 2003). On the other hand, cooling
triggers catastrophic instabilities unless closely restrained by
other processes feeding energy back to baryons (see Blanchard,
Valls-Gabaud \& Mamon 1992); so energy \emph{additions} $\Delta E
>0$ are mandatory anyway, and will constitute our focus next.

\begin{figure}
\epsscale{1.0}\plotone{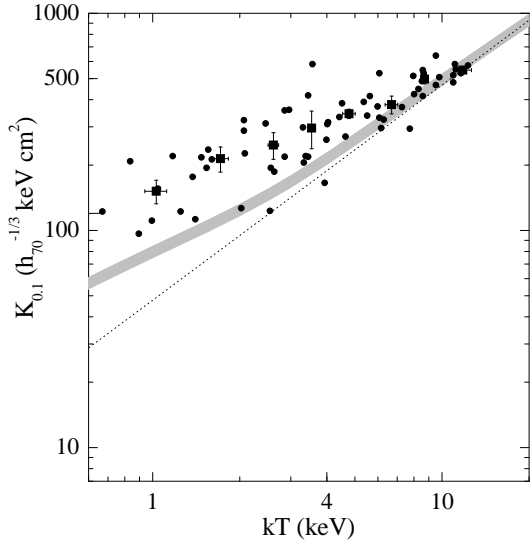}\caption{Central entropy
$K_{0.1}$ (at $r\approx 0.1\, R$) vs. X-ray temperature $T$. Data
for clusters and groups are from Ponman et al. (2003):
\textit{circles} mark individual systems and \textit{squares}
refer to binned data. The \textit{dotted line} represents pure
gravitational heating. The \textit{strip} (with 2 $\sigma$ width
provided by the merging histories) illustrates our results for SN
preheating with $k\Delta T = \case{1}{4}$ keV per particle, as
discussed in \S~3.}
\end{figure}

%
\section{The need for AGN feedback}
%
Obvious energy sources are provided by Type II SN explosions; do
they contribute enough energy feedback? SNe provide energies
$E_{SN}\approx 10^{51}$ ergs with an occurrence $\eta_{SN}\lesssim
5\times 10^{-3}$ per solar mass condensed into stars, the latter
value being calibrated so as to include the yield of strong winds
from young hot stars (Bressan, Chiosi \& Fagotto 1994).

Such outputs may be coupled to the surrounding gas at levels
$f_{SN}\la \case{1}{2}$ when cooperative SN remnants propagation
takes place, as in the case of starbursts, to drive subsonic
galactic winds (see Matteucci 2003). Then the integrated thermal
input attains the maximal level (Cavaliere, Lapi \& Menci 2002)
\begin{equation}
k \Delta T = \frac{2\, \mu m_p}{3}\, f_{SN}\, E_{SN}\, \eta_{SN}\,
{m_{\star}\over m} \lesssim \frac{1}{4}~ \mathrm{keV~
particle^{-1}}~
\end{equation}
in groups with stellar to gas mass ratios up to $m_{\star}/m
\approx \case{1}{2}$ (see David 1997); somewhat smaller values
obtain in clusters (e.g., Lin, Mohr \& Stanford 2003).

The SN feedback and the originating star formation are described
by semianalytic models, in particular that of Menci \& Cavaliere
(2000). They base their model on the DM merging histories, i.e.,
the hierarchical buildup of a galaxy or a group with their DM and
baryonic contents through merging events with comparable or
smaller partners, down to nearly smooth inflow (Lacey \& Cole
1993). The model, in addition, specifies how the baryons are
cycled between the cool, stellar, and hot phases; the latter
contributes to the ICP/IGP, while the former two phases yield the
stellar observables.

It is found that most of a structure's DM mass $M$ (and of the IGP
mass likewise) is contributed to its main progenitor by merging
partners with masses $M'\la M/3$ and related virial temperatures
$T'_v\la 0.6\, T_v$ (Cavaliere, Menci \& Tozzi 1999). The smaller
lumps have shallower gravitational wells and produce more
star-related energy on scales closer to their dynamical times; so
they are more effective in heating up their gas share to
temperatures $T_v'+\Delta T$. During each subsequent step of the
hierarchy forming larger groups or clusters, the {\it externally}
preheated gas (see Muanwong et al. 2002) will be hindered from
flowing in and contributing to the IGP or ICP. So under any model
depleted densities will be propagated some steps \emph{up} the
hierarchy.

In sum, SNe make optimal use of their energy in {\it preheating}
the IGP. However, their input $k\Delta T\approx \case{1}{4}$ keV
per particle turns out to cause only limited luminosity
depressions or entropy enhancements, as shown by the light strips
in Figs.~1 and 2. The result may be understood by referring to the
simple isothermal case where $L_X\propto n_2^2\int{dr}\, r^2\,
e^{2\beta\Delta\phi}$ applies; in moving from rich to poor
clusters $n$ decreases, governed mainly by the decreasing
exponential $e^{2\, \beta\, \Delta\phi}$ (as visualized by Fig.~5
of Cavaliere \& Lapi 2005). But the normalized DM potential
$\Delta\phi$ deepens because of the increased concentration (an
intrinsic feature discussed in Appendix A). To offset this trend
and provide constant or decreasing density, it is clearly required
that $\beta\approx T_v/(T_v+\Delta T)$ be lowered from the cluster
value by a sufficiently strong preheating $\Delta T$; in detail,
the approximation $\beta\approx 0.7 - \Delta T/T$ holds; see
Appendix B and in particular Eq.~(B9) with $\theta=T/\Delta T$.
Numerically, the requirement comes to $k\Delta T > 0.5$ keV per
particle for any significant luminosity depression in a poor
cluster; yet more preheating is required with polytropic plasma
distributions. On the other hand, including Type I\textit{a} SNe
still does not meet the above requirement (see Pipino et al.
2003).

In view of these SN limitations, in the rest of the paper we
concentrate on the stronger feedback provided by quasars and
active galactic nuclei (see Valageas \& Silk 1999; Wu, Fabian \&
Nulsen 2000; Yamada \& Fujita 2001; Nath \& Roychowdhury 2002).
These sources are kindled when sizeable amounts of galactic gas,
triggered by mergers or interactions of the host with companion
galaxies, are funneled downward from kpc scales; they not only
form circumnuclear starbursts but eventually trickle farther down
to the very nucleus (see Menci et al. 2004) and accrete onto a
central supermassive BH.

%
\section{External preheating from AGN\lowercase{s}}
%

On accreting the BH mass $M_{\bullet}$, the integrated energy input
to the surrounding plasma comes to values
\begin{equation}
k\Delta T = {2\, \mu m_p\over 3}\, f\, \eta\,  c^2\,
{M_{\bullet}\over 4\, M_{b}}\, {m_{\star}\over m}\approx {1\over
2}~\mathrm{keV~particle^{-1}}~,
\end{equation}
easily larger than for SNe. We have used the standard mass-energy
conversion efficiency $\eta\approx 10^{-1}$, and the locally
observed ratio $M_{\bullet}/M_{b}\approx 2\times 10^{-3}$ of BH to
galactic bulge masses (Merritt \& Ferrarese 2001); the factor
$\case{1}{4}$ accounts for the bulge mass observed in blue light
compared to that integrated over the star formation history
(Fabian 2004a). Finally, we will adopt values $f\approx 5\times
10^{-2}$ for the fractional AGN output actually coupled to the
surrounding gas, on the grounds discussed next.

The 10\% radio-loud AGNs directly produce considerable kinetic or
thermal energies in the form of bubbles and jets (see Forman et
al. 2004), but statistics and nonspherical geometry reduce their
average contribution to $f$. On the other hand, in the 90\%
radio-quiet AGNs a small coupling is expected for the radiative
output because of the flat spectrum and low photon momenta; the
observations of wind speeds up to $v_w\approx 0.4\, c$ suggest
values around $v_w/2\, c\approx 10^{-1}$ associated with covering
factors of order $10^{-1}$ (see Chartas, Brandt \& Gallagher 2003;
Pounds et al. 2003). We shall see that average values around
$f\approx 5\times 10^{-2}$ are consistent not only with the X-ray
observations of the IGP, but also with the mainly optical
observations of the relic BHs in galaxies.

Considering that the AGN activity closely parallels the star
formation in spheroids (Franceschini et al. 1999; Granato et al.
2004; Umemura 2004), we add the AGN energy injections to SN's to
obtain preheating energies up to $k\Delta T\approx \case{3}{4}$
keV per particle. Such a combined value produces a sizeable step
toward the locus of the data, as shown in Figs.~3 and 4 by the
heavy strips. How we obtain these is explained next.

Toward this purpose it is convenient to discuss further the
\emph{modus operandi} of the external preheating. During the
formation of a DM structure, outer lumps and the associated gas flow
in together; but just inside $R$ the smaller and/or less bound gas
bunches are promptly stripped away from their DM hosts, while
gaining entropy (Tormen, Moscardini \& Yoshida 2004, see their
Fig.~8). The outcome constitutes a complex patchwork of shocks of
all sizes, comprised within an outer \emph{layer} with thickness
$\delta\lesssim 10^{-1}\, R$ wherein most of the entropy rise takes
place.

The net result is close to that computed from considering a coherent
accretion shock, roughly spherical and located at $r\approx R$ as
considered by Cavaliere et al. (1999), Dos Santos \& Dor\'e (2002),
and Voit et al. (2003). In fact, across the layer we may retrace the
classic Rankine-Hugoniot derivation based on the conservation of
mass, momentum and energy for plasma particles with $3$ degrees of
freedom; for a reasonably thin layer we recover the standard entropy
jump across a shock
\begin{equation}
K_2 = K_1\, \theta^{5/3}\, \left[2\, \left(\theta-1\right) +
\sqrt{4\, \left(\theta-1\right)^2 +\theta}\right]^{-2/3}~
\end{equation}
to within $\mathcal{O}(\delta/R)$ accuracy, \emph{independently}
of geometrical details (as discussed in Appendix B). To within the
same accuracy, the strength parameter $\theta\equiv T_2/T_1$,
i.e., the ratio of the down- to the upstream temperature is linked
by
\begin{equation}
\theta = {5\, \mathcal{M}^2\over 16} + {7\over 8} - {3\over 16\,
\mathcal{M}^2}~
\end{equation}
to the Mach number $\mathcal{M}=(3\, \mu m_p\, \tilde{v}^2_1/5\,
kT_1)^{1/2}$ of the flow velocity $\tilde{v}_1$ relative to the
shock/layer. Henceforth we refer to ``shocks'', but we also
include thin layers.

We now consider in closer detail the \emph{combined} preheating by
SNe and AGNs; this comes into play through $T_1=T_v'+\Delta T$
that enters $K_1$ and $\theta$ in Eqs.~(5) and (6). We average the
resulting $K_2$ over the full structure's merging history that
includes the distributions of progenitor masses $M'$ or related
$T_v'$; to this purpose we implement, as in Cavaliere et al.
(1999), the conditional probabilities and the merging rates from
the standard cold DM cosmogony as given by Lacey \& Cole (1993).
This straightforward if laborious procedure (which is dominated by
the smaller partners and so further validates Eqs.~[5] and [6], as
discussed in Appendix B) is made semianalytically, and yields the
heavy strips in Figs.~3 and 4. Their width illustrates the
variance (at $95\%$ probability level) around the mean value,
induced mainly by the merging stochasticity; the smooth, low-power
AGN activity considered here does not contribute much additional
scatter.

\begin{figure}
\epsscale{1.0}\plotone{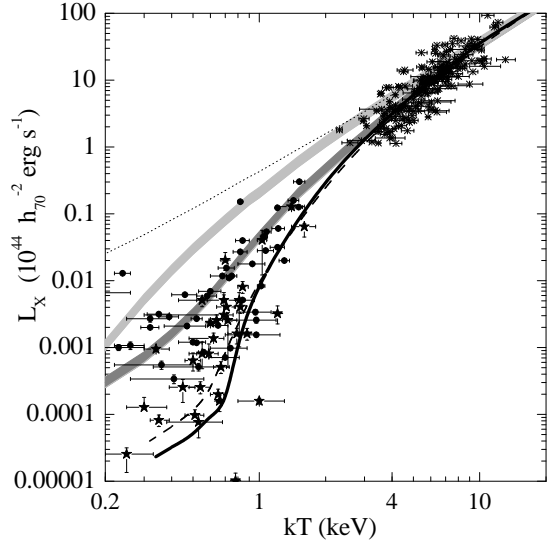}\caption{Integrated X-ray
luminosity $L_X$ vs. X-ray temperature $T$. Data, \textit{dotted
line} and \textit{light shaded strip} as in Fig.~1. The
\textit{heavy shaded strip} (with 2 $\sigma$ width provided by the
merging histories) illustrates our results for external preheating
when including the AGN contribution to a total $k\Delta T =
\case{3}{4}$ keV per particle, as discussed in \S~4. Our results
for the internal impacts from quasars are illustrated by the
\textit{solid} (ejection model) and \textit{dashed} (outflow
model) lines, see \S~5 for details. The coupling level of the
quasar output to the ambient medium is $f = 5\times 10^{-2}$.}
\end{figure}

Next we discuss why the results fit clusters better than groups.
In very {\it rich} clusters the infall velocities $v_1\approx
2.1\, (kT_v/\mu m_p)^{1/2}$ are large, and $\tilde{v}_1\simeq 4\,
v_1/3$ is larger yet (see Appendix B). These velocities are
dominantly supersonic, except for the few major lumps which carry
warm gas deep into the structure (as observed by Mazzotta et al.
2002) and contribute little to prompt entropy gains. So the
effective shocks are uniformly \emph{strong} with $\theta\simeq
\mu m_p\, v_1^2/3\, kT_1\gg 1$, see Eq.~(B7). Such conditions in
Eq.~(5) yield $K_2/K_1 \simeq \theta/4^{2/3}$, corresponding to
nearly constant $n_2/n_1\approx 4$, see Eq.~(B7); they also yield
a nearly constant value of $\beta=T_v/\theta\, T_1\simeq 3\,
kT_v/\mu m_p\, v_1^2\approx \case{2}{3}$. In other words, here we
find pure gravitational heating at work to enforce $K\propto T$ or
$L_X \propto T^2$.

The related, raising entropy profiles reflect the history of
progressive depositions of shells undergoing stronger and stronger
shocks during the hierarchical growth to a rich cluster; in fact,
in the outer regions we find $\kappa\propto r^{1.1}$ and
$\Gamma\approx 1.1$. We derive these values by reducing to bare
bones the model of Tozzi \& Norman (2001). We adopt in full the
Concordance Cosmology, and nearly self-similar hierarchical
collapse with a constant perturbation power index around $-1.2$
appropriate for rich clusters (see Padmanabhan 2003); these
conditions imply the last accreted shell to add a mass $\Delta
M\propto M$ on top of the mass $M\propto (1+z)^{-3.2}$ virialized
at $z\ga \case{1}{2}$ (see also Lapi 2004). In the process, the
entropy $K\propto T/n^{2/3}$ grows because the strong shocks
prevailing in rich clusters yield not only $T\approx T_v$ and
$n_2\simeq 4\, n_1$ as above, but also $n_1\propto \rho\propto
(1+z)^{2.9}$ considering the appropriate collapse threshold; in
terms of $m\propto M$ this translates into $\kappa\propto
m^{2/3}\, (1+z)^{-1}\propto m$.

For the ICP in equilibrium, the radial entropy profile $\kappa(r)$
corresponding to this distribution $\kappa(m)$ is found as
follows. In the outskirts we approximate the entropy profile as
$\kappa(r)\propto r^{\alpha}$, with $\alpha$ to be determined;
then Eq.~(A3) implies $n(r)\propto r^{-3\, (\alpha + 2)/5}$, and
we obtain $\kappa(r)\propto m(<r)^{5\, \alpha/3\,(3-\alpha)}$ on
considering that $m(<r)=4\pi\, m_p\, \int^r{dx}\, x^2\, n(x)$
holds. Requiring consistency with the entropy distribution
$\kappa\propto m$ derived above, we obtain $\alpha\approx 1.1$. So
our final results read $\kappa(r)\propto r^{1.1}$ and $n(r)\propto
r^{-1.9}$, which accord with the data by Ponman, Sanderson \&
Finoguenov (2003), and with the simulations by Tornatore et al.
(2003); the related value $\Gamma\approx 1.1$ agrees with that
observed by Ettori \& Fabian (1999) and De Grandi \& Molendi
(2002).

\begin{figure}
\epsscale{1.0}\plotone{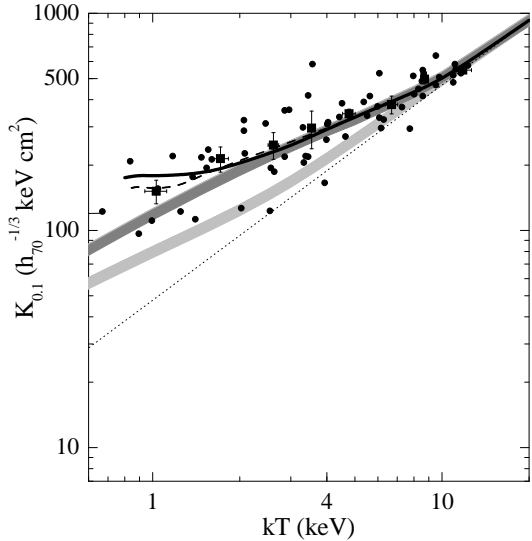}\caption{Central entropy
$K_{0.1}$ vs. X-ray temperature $T$. Data, \textit{dotted line}
and \textit{light shaded strip} as in Fig.~2. The \textit{heavy
shaded strip} (with 2 $\sigma$ width provided by the merging
histories) illustrates our results for external preheating when
including the AGN contribution to a total $k\Delta T =
\case{3}{4}$ keV per particle, as discussed in \S~4. Our results
for the internal impacts from quasars are illustrated by the
\textit{solid} (ejection model) and \textit{dashed} (outflow
model) lines, see \S~5 for details. As before, the coupling level
is $f = 5\times 10^{-2}$.}
\end{figure}

In {\it poor} clusters, on the other hand, the infall is slower
with $\tilde{v}_1\simeq 4\, v_1/3 + 5\, kT_1/4\, \mu m_p\, v_1$,
see Appendix B. Now the inflow is \emph{less} supersonic, and the
accretion shocks are easily modulated by the preheating
temperature to a strength $\theta\simeq \mu m_p\, v_1^2/3\, kT_1 +
3/2$, see Eq.~(B8). Less entropy is produced by these intermediate
shocks, while an additional contribution is just carried in with
the warm inflowing gas, to yield $K_2/K_1 \simeq (\theta+
5/8)/4^{2/3}$; correspondingly, the boundary densities are lowered
to $n_2/n_1\simeq 4 - 15/4\, \theta$, see Eq.~(B8). In addition,
the density profiles are now just flatter than in rich clusters,
since $n(r)/n_2$ is appreciably decreased with preheating levels
$k\Delta T\approx \case{3}{4}$ keV per particle that satisfy the
condition derived at the end of \S~3; these are effective in
lowering all densities, hence in depressing $L_X$ and enhancing
$K$.

In \emph{groups} and galaxies these preheating levels are enough to
cause smoother, transonic inflows and \emph{weak} shocks with
$\theta\simeq 1$, yielding small jumps $K_2/K_1\simeq 1+5\,
(\theta-1)^3/6$ and $n_2/n_1\simeq 1$, see Eq.~(B10). The X-ray
scaling relations produced by the combined external preheating of
SNe and AGNs (heavy strips in Figs. 4 and 3) are in marginal
agreement with the trends in the data, while the wide scatter is
still unaccounted for. Moreover, the weak shocks so produced would
imply nearly flat profiles $\kappa(r)$, which often are not observed
(Pratt \& Arnaud 2003; Rasmussen \& Ponman 2004).

The problem with such isentropic profiles would be aggravated and
propagated to poor clusters while the entropy would be raised too
much, if one tuned high the AGN preheating, much above the level
$\case{1}{2}$ keV per particle given by Eq.~(4). From the previous
relations the problem is easily seen to develop even before
solving the luminosity issue in groups.

\begin{figure}[b]
\epsscale{1.0}\plotone{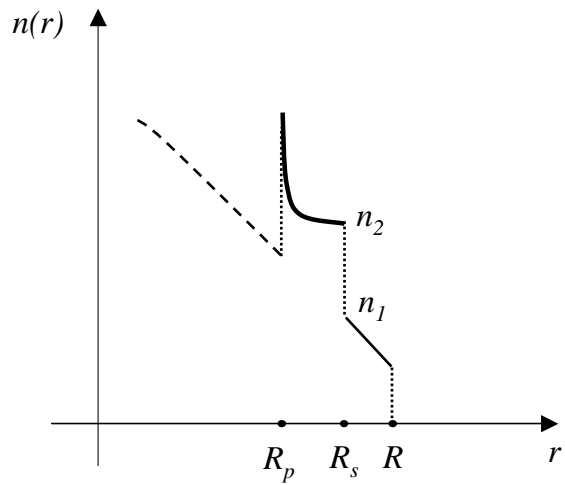}\caption{An outline of the
density distribution during the propagation of a quasar-driven
blast throughout the equilibrium plasma. By the \textit{dashed}
line we represent the initial density run $n\propto r^{-2}$ in the
volume already evacuated by the blast, by the \textit{thick solid}
line the perturbed density in the blast (specified in Fig.~C1),
and by the \textit{thin solid} line the still unperturbed density.
The perturbed flow is confined between the trailing piston at
$R_p$ and the leading shock at $R_s$; in our illustration this is
still far from the virial radius $R$.}
\end{figure}

%
\section{Internal impacts of quasars}
%
But right in groups and galaxies the impulsive inputs by powerful
quasars take over, providing from \emph{inside} an additional impact
on the IGP that can cause outflow or ejection. For this to occur,
two energies compete: the overall input $\Delta E\approx 2\times
10^{62}\, f\,$ $(M_{\bullet}/10^9\, M_{\odot})\, (1+z)^{-3/2}$ ergs
provided by a quasar on accreting the mass $M_{\bullet}$ within the
host dynamical time $t_d\approx 10^8$ yr set by mergers or
interactions; the (absolute) total energy $E\approx 2\times
10^{61}\, (kT/\mathrm{keV})^{5/2}\, (1+z)^{-3/2}$ ergs residing in
the equilibrium IGP (Lapi, Cavaliere, \& De Zotti 2003).

The relevant ratio
\begin{equation}
{\Delta E\over E}\approx 0.5\, {f\over
5\times 10^{-2}}\, {M_{\bullet}\over 10^9\, M_{\odot}}\,
\left({kT\over \mathrm{keV}}\right)^{-5/2}
\end{equation}
is small in clusters but increases toward groups, and approaches
unity in poor groups with $kT\approx 1$ keV to attain a few in
large galaxies with $kT\approx \case{1}{2}$ keV. Within the
central kpc of such structures the quasar launches a piston (see
King 2003; Granato et al. 2004) that drives through the
surrounding plasma a blast wave bounded by a leading shock at
$r=R_s$ (see Fig.~5). These blasts constitute effective,
quasi-isotropic means to propagate energy far away from the
central source.

While the latter shines, the blast affects the plasma out to the
distance $R_s$ where the initial energy $E(<R_s)$ is comparable to
the cumulative input $\Delta E(t)$. In fact, the condition $\Delta
E(t)/E(<R_s)=$ const defines the self-similar propagation of the
blast and the motion $R_s(t)$ of the leading shock; as shown in
Appendix C, the result is $R_s(t)\propto t^{2/\omega}$ in an
initial distribution $n(r)\propto r^{-\omega}$ ($2\leq \omega <
2.5$) for the plasma density under the energy input $\Delta
E(t)\propto t^{2\, (5-2\, \omega)/\omega}$. The simplest flow
obtains with $\omega=2$ (the standard isothermal sphere) implying
a source power $L(t)\equiv d\Delta E/dt=$ const; when $\omega>2$
applies the power $L(t)\propto t^{5\, (2-\omega)/\omega}$
declines, a useful means to describe the quasar fading because of
its own feedback on the accreting gas.

This new family of hydrodynamic solutions, proposed by Cavaliere
et al. (2002), is described in Appendix C; it is used below to
evaluate mass loss from, and entropy distribution into, the
structures. The solutions include the restraints set to
gasdynamics by a nonzero initial pressure $p(r)\propto
r^{2\,(1-\omega)}$ and by the DM gravity; so not only do they
imply a well-defined $E(<R_s)$, but they also cover the full range
of blast strengths from weak in clusters to strong in galaxies,
depending on the magnitude of the key parameter $\Delta E/E$.

As $\Delta E/E$ increases, so does the Mach number $\mathcal{M}$
of the leading shock; their relation is shown in Fig.~C2 and is
approximated by $\mathcal{M}^2\simeq 1+ (\Delta E/E)$ for $\Delta
E/E\la 2$ in the simple $\omega=2$ model. Meanwhile, the ratio of
the kinetic to the thermal energy ranges up to $2$, see Eq.~(C17).
Correspondingly, within the time $t_d$ increasing plasma amounts
are driven beyond the virial radius $R$ of a large galaxy or a
poor group; in the simple $\omega=2$ model the fractional mass
ejected or flowed out is well approximated by $\Delta m/m\simeq
\Delta E/2\, E$, see also Table~1.

These results turn out to be nearly independent of the specific
mode for mass loss; in particular, we compare two extreme cases.
In the first one (``ejection''), we take $\Delta m$ to be the mass
in the blast driven outside of the virial radius $R$ at $t=t_d$ by
the blast kinetic energy. In the second case (``outflow''), we
adopt constant pressure as boundary condition at $r=R$, to obtain
new densities $n'\propto (T+\Delta T)^{-1}$; now $\Delta m$ is the
mass flowed out of the structure due to the extrathermal energy
$\Delta T$ deposited by the blast. Beyond model details, we find
that the mass loss closely obeys $\Delta m/m\simeq \Delta E/2\,
E$.

In both cases, after the passage of the blast the IGP will recover
hydrostatic equilibrium, described by Eq.~(2). But all new
densities $n'$ will be \emph{depleted} by the factor $1-\Delta
m/m$ below the initial value already affected by the preheating
from SNe and AGNs. In addition, in both cases the extrathermal
energy deposited by the blast lowers the values of $\beta'$; this
is given in Table~1, and may be understood in terms of
$\beta'/\beta\approx T/(T+\Delta T)$. The resulting $L_X\propto
(1-\Delta m/m)^2$, including the appropriate $T$-dependent shape
factor, is shown by the solid and dashed lines in Fig.~3 for both
models, ejection and outflow, discussed above.

The IGP entropy is increased by quasar-driven blasts. While these
sweep through the plasma, a moderate production takes place across
the leading shock, and leads to a jump $K_2/K_1$ given again by the
general Eqs.~(5) and (6). In the equilibrium recovered after the
plasma mass loss $\Delta m/m$ caused by ejection or outflow, the
entropy is further enhanced to read
\begin{equation}
K'_2/K_2 = \left(1-\Delta m/ m\right)^{-2/3}~.
\end{equation}
For example, in a group with $kT = \case{3}{4}$ keV the combined
preheating by SNe and AGNs yields entropy levels corresponding to
$100$ keV cm$^2$. This is raised to $180$ keV cm$^2$ by the
internal blast driven by a quasar deriving from a BH of $10^{9}\,
M_{\odot}$ with coupling $f\approx 5\times 10^{-2}$, that produce
$\Delta E/E\approx 1$. The resulting central entropy $K_{0.1}$
including the appropriate $T$-dependent shape factor is shown by
the solid and dashed lines in Fig.~4 for both models, ejection and
outflow, discussed above.

\begin{deluxetable}{lcccc}[b]
\tabletypesize{} \tablecaption{Relevant Blast Wave Quantities
($\omega=2$)} \tablewidth{0pt} \tablehead{\colhead{$\Delta E/E$} &
\colhead{$\mathcal{M}$} & \colhead{$\langle p\rangle/p_1$} &
\colhead{$1-\Delta m /m$} &\colhead{$\beta'/\beta$}}\startdata
0.3......  & 1.2 & 3.6 &  0.92      & 0.94 \\
1......... & 1.5 & 4.6 &  0.58      & 0.86\\
3......... & 1.9 & 6.3 &  $\sim 0$  & - \\
\enddata
\end{deluxetable}

Relatedly, the entropy profiles are \emph{steep} after the blast
passage. They arise when the entropy produced in the blast, clearly
piled up toward the leading shock, is redistributed in the recovered
equilibrium. We find
\begin{equation}
\kappa(m)\propto m^{4/3}
\end{equation}
to hold in terms of the plasma mass $m$ swept up by the blast (see
Fig.~C1 and Eq.~[C11]). In the adiabatically recovered equilibrium
we require the entropy distribution $\kappa'(m)$ to equal
$\kappa(m)\propto m^{4/3}$, and proceed in analogy with the
technical steps used in \S~3.

In detail, let us approximate $\kappa'\propto r^{\alpha}$, with
$\alpha$ to be determined; then Eq.~(A3) implies $n'\propto
r^{-3\, (\alpha + 2)/5}$, and thus $\kappa'\propto m^{5\,
\alpha/3\,(3-\alpha)}$ follows. Requiring this to be consistent
with the entropy distribution $\kappa\propto m^{4/3}$ in the blast
(as anticipated above) yields $\alpha\approx 1.3$. In other words,
the blast acting from inside leaves in the readjusted plasma a
strong imprint of its own entropy distribution, in the form of a
\emph{steep} profile $\kappa'(r) \propto r^{1.3}$ consistent with
the data.

%
\section{Discussion and conclusions}
%
In this paper we have used pilot semianalytic modeling to show
that the energy fed back to baryons by AGNs and quasars is
essential to fit the recent X-ray data. We find that the AGN
external preheating dominates over SN's to yield the scaling laws
$L_X\propto T^3$ and $K\propto T^{2/3}$ related by Eq.~(1), that
constitute fitting trends in clusters if still marginal for
groups. But in groups and in large galaxies the quasar impulsive
feedback acting from inside takes over, to eject some plasma and
further \emph{depress} $L_X$, while \emph{enhancing} $K$ and
originating \emph{non-isentropic} profiles.

We stress that the energy added by AGNs plays an \emph{inverse} role
in preheating and in ejection/outflow. This is because in moving
from clusters to groups the ratio $\Delta E/E$ of the added vs. the
equilibrium energy is bound to increase. This causes relatively
higher external preheating, warmer inflows and \emph{weaker}
accretion shocks; on the other hand, it drives \emph{stronger}
internal blasts causing more mass ejection/outflow. In parallel, the
leading shocks of the blasts replace the accretion shocks in the
role of increasing the outer entropy production, so originating
comparably steep entropy profiles.

Such internal effects, however, are to \emph{saturate} in galaxies
because the values of $\Delta E/E$ there are limited, lest the
impulsive quasar feedback ejects so much of the surrounding
baryons as to stop the BH accretion altogether (see Silk \& Rees
1998); the saturation is what in our calculation yields the lower
\emph{elbows} of the solid and dashed lines presented in Fig.~3.
As argued in Cavaliere et al. (2002), the system constituted by
the BH and the surrounding baryons self-regulates to the verge of
unbinding; the condition $\Delta E\approx E$ directly yields
$M_{\bullet}\approx$ $5\times 10^8\, M_{\odot}\, (f/5\times
10^{-2})^{-1}\, (\sigma/ 300\, \mathrm{km\, s^{-1}})^5$ in terms
of the DM velocity dispersion $\sigma$. In turn, the latter is
found to correlate less than linearly with the velocity dispersion
$\sigma_{\star}$ of the host galactic bulge (Ferrarese 2002; Baes
et al. 2003; A. Pizzella et al. 2005, in preparation); so the BH
mass approaches $M_{\bullet}\propto \sigma_{\star}^4$.

Thus for the \emph{same} value $f\approx 5\times 10^{-2}$
indicated by the average X-ray data for groups we agree (within
the observed scatter, see Tremaine et al. 2002) with the
observations of the uppermost relic BH masses in the bulges of
many local and currently inactive galaxies. Similar BH masses may
be also contributed by an initial, supercritical accretion phase
(as discussed by King 2003) launching the piston that in turn
drives the far-reaching blasts described above. Subsequently, our
outgoing self-similar blasts with $\Delta E(t)/E(<R_s)\approx$
const stay tuned to the condition $\Delta E\approx E$.
Specifically, for $\omega\rightarrow 2.5$ not only $E(<R_s)\propto
R_s^{5-2\, \omega}\rightarrow$ const holds but also $\Delta
E\propto t^{2\, (5-2\, \omega)/\omega}\rightarrow$ const applies,
consistent with fading quasar output; thus at most limited
increase of BH mass $M_{\bullet}$ takes place.

The empirical fact (see Wandel 2002; Vestergaard 2004) that a
similar $M_{\bullet}-\sigma$ correlation appears to hold also for
the currently active and faraway BHs is consistent with our adoption
of comparable values for $M_{\bullet}$ energizing both modes of
nuclear activity: the impulsive quasar feedback effective for plasma
ejection from groups and galaxies, and the smoother, long-lived AGN
outputs sufficient to preheat the gas falling into poor clusters. We
are also consistent with the rough equality of the mass densities in
BHs derived from the powerful emissions of the quasars in the
optical band (see Marconi et al. 2004), and from the weaker and
later AGN activity detected mainly in X-rays (Hasinger 2004; Fabian
2004a).

On the other hand, strong and rare (i.e., increasingly sporadic)
quasar impacts can also explain the \emph{scatter} of the X-ray
data widening toward smaller systems as poor groups or massive
galaxies (Mushotzky 2004). This we trace back to the increasing
\emph{variance} in the occurrence of strong quasar events or even
in their coupling level $f$, that concur to dynamically modulate
the plasma ejection $\Delta m/m\propto f\, M_{\bullet}$ and
nonlinearly affect $L_X\propto (1-\Delta m/m)^2$ when $\Delta m/m$
approaches $1$. As the hierarchical clustering proceeds toward
clusters, instead, the evolution of the quasars cuts down most
internal effects; this is because the impulsive contributions to
$\Delta E$ within a structure's dynamical time hardly can keep
pace with the increase of the equilibrium energy $E\propto m\,
T_v$ in such late and massive systems with deep potential wells.

We stress that our upper and lower bounds for $L_X$ illustrated in
Fig.~3 by the SN strip and the quasar line comprise nearly
\emph{all} data points, except for a few groups with peculiar
features currently under scrutiny (see Mushotzky 2004; Osmond \&
Ponman 2004). So we submit that several pieces of data fit
together when considering both the external preheating from AGNs
and the internal impact from quasars, with the same average values
of $f\, M_{\bullet}$. We remark that several authors (e.g.,
Ruszkowski \& Begelman 2002; Fabian 2004b; Zanni et al. 2004) have
argued the relevance of AGN feedback in explaining the puzzle
posed by the ``cool cores'' (Molendi \& Pizzolato 2001) at the
very centers of many clusters. On the other hand, Cavaliere et al.
(2002) and Granato et al. (2004) have stressed that powerful
quasar impacts in massive spheroids easily quench star formation
and produce precociously red giant elliptical galaxies.

To conclude, we stress that energy feedback from AGNs and quasars
with an overall coupling around $5\times 10^{-2}$ to the ambient
baryons yields agreement with independent observations in
different frequency bands and over different distance scales.
Specifically, this paper is focused on the extended X-ray
emissions and plasma entropy of poor clusters and groups; but we
have also considered at galactic and subgalactic scales the mainly
optical correlation of nuclear BH masses vs. host velocity
dispersions. At the intermediate scales of early massive galaxies
and in the $\mu$wave/submm band, we have proposed in Lapi et al.
(2003) how to catch quasar impacts in the act from resolved
Sunyaev-Zel'dovich signals enhanced by overpressure in running
blast waves. Such transient events sweeping plasma outward to
lower densities (see Figs.~5 and C1) hardly increase the extended
X-ray emissions; they instead specifically correlate with
pointlike X-rays from a fully active quasar, and/or with strong IR
emissions signaling a nascent quasar enshrouded by dust.

\begin{acknowledgements}
We acknowledge fruitful discussions with G. Tormen, and the timely
and helpful comments by our referee. This work is partially
supported by INAF and MIUR grants.
\end{acknowledgements}

%

\begin{appendix}

\section{Hydrostatic equilibrium}

The hot plasma constituting the ICP/IGP pervades the potential wells
of clusters and groups, being in overall virial equilibrium with the
DM. As the sound crossing time is comparable to, or somewhat shorter
than the structure dynamical time, hydrostatic equilibrium applies;
when the thermal pressure $p=n\, k T/\mu$ is dominant (see Ricker \&
Sarazin 2001; Inogamov \& Sunyaev 2003) this yields
\begin{equation}
{1\over m_p\, n}\, {d p\over d r} = - {d \Phi \over d r}~,
\end{equation}
in terms of plasma number density $n$, and of the temperature $T$.
The solution of this differential equation requires one
\emph{boundary} condition, for example the value $n(R)=n_2$ at the
virial radius $r=R$; it also requires an equation of state, i.e., a
specific relation between $n(r)$ and $T(r)$.

As to the DM potential $\Phi(r)$, we adopt the widely used NFW form
(Navarro, Frenk \& White 1997)
\begin{equation}
\Phi(r) \approx - 3\, \sigma^2\, g(c)\, {\ln{(1+r/r_s)}\over r/r_s}~,
\end{equation}
involving the one-dimensional velocity dispersion $\sigma$ of the
DM, and the scale $r_s \equiv R/c$. The concentration parameter
$c\approx 5\, (M/10^{15}\, M_{\odot})^{-0.13}$ slowly increases
(Bullock et al. 2001), and the factor $g(c) =[\ln{(1+c)} -
c/(1+c)]^{-1}$ weakly rises from clusters to groups. To wit, the
smaller, earlier DM halos are more concentrated in terms of the
normalized potential $\phi\equiv\Phi/\sigma^2$, consistent with
the tenets of hierarchical structure formation (see Padmanabhan
2003).

It is useful to recast the hydrostatic equilibrium in terms of the
all-important adiabat $K=kT/n^{2/3}$ to obtain
\begin{equation}
{5\over 3}\, {d\ln n\over d\ln r} + {d\ln K\over d\ln r}= -
{kT_v\over K\, n^{2/3}}\, {d \phi \over d\ln r}~;
\end{equation}
recall that $kT_v=\mu m_p\sigma^2$ is the virial temperature. The
above is a first-order differential equation of the Euler type for
$n(r)$; in terms of $K(r)$, it can be formally integrated by
standard methods to obtain Eq.~(2).

In the cluster outskirts, Eq.~(A3) directly relates the slopes of
$n(r)$ and $K(r)$, so that $n(r)\propto r^{-3\, (\alpha+2)/5}$
corresponds to $K(r)\propto r^{\alpha}$. This is easily derived
near $r\approx R$, where for the NFW potential $d\phi/d\ln
r\approx 3$ holds; meanwhile the coefficient $T_v/K\,
n^{2/3}\approx T_v/T_2$ is easily recognized to be $\beta$, with
values close to $\case{2}{3}$ for clusters.

\section{Accretion shocks and layers}

Across any sharp transition like a shock, the conservation laws of
mass, total stress, and energy for a plasma with $3$ kinetic
degrees of freedom are written (Landau \& Lifshitz 1959)
\begin{eqnarray}
\nonumber& & n_1\, \tilde{v}_1  =  n_2\, \tilde{v}_2\\
\nonumber\\
\nonumber& & p_1 + m_p\, n_1\, \tilde{v}_1^2  =  p_2 + m_p\, n_2\,
\tilde{v}_2^2\\
\nonumber\\
\nonumber& & {1\over 2}\, m_p\, n_1\, \tilde{v}_1^3 + {5\over 2}\,
p_1\, \tilde{v}_1 = {1\over 2}\, m_p\, n_2\, \tilde{v}_2^3 + {5\over 2}\,
p_2\, \tilde{v}_2~.\\
\end{eqnarray}
As is customary, we have indicated with the subscripts $1$ and $2$
the pre- and postshock variables, respectively; in addition, by
$\tilde{v}$ we indicate velocities measured in the shock
\emph{rest} frame.

The previous system of equations leads after some algebra to the
temperature jump $T_2/T_1\equiv \theta$ under the \emph{general}
form
\begin{equation}
\theta = {5 \over 16}\, \mathcal{M}^2+{7\over 8}-{3\over 16}\,
\frac{1}{\mathcal{M}^2}~,
\end{equation}
in terms of the Mach number $\mathcal{M}\equiv (3\, \mu m_p\,
\tilde{v}_1^2/5\, k T_1)^{1/2}$;  this is  Eq.~(6) of the main text.
It is seen that shock heating ($\theta > 1$) requires the flow to be
supersonic in the shock rest frame, i.e., $\mathcal{M}> 1$ as
expected. The corresponding density jump reads
\begin{equation}
{n_2\over n_1} = \frac{4\, \mathcal{M}^2}{\mathcal{M}^2+ 3}~;
\end{equation}
or in terms of $\theta$ (cf. Cavaliere et al. 1999)
\begin{equation}
{n_2\over n_1} = 2\, \left( 1-{1\over \theta}\right) + \sqrt{4\,
\left( 1-{1\over \theta} \right)^2 +{1\over \theta}}~.
\end{equation}
This may be used to express the postshock adiabat $K_2=k T_2/
n_2^{2/3}$, which leads to Eq.~(5) of \S~4 in the context of
accretion flows. The same general equation applies also to the
leading shock of an outgoing blast wave, discussed in \S~5.

Focusing now on \emph{accretion} flows, we consider what happens
if the transition occurs across a layer of finite thickness
$\delta$ located at $r\approx R$. In such a case, the conservation
equations above include additional terms due to volume forces or
nonplanar geometry; e.g., in the momentum equation the
gravitational term $\int_{R-\delta}^R{dr}\, G\, M(<r)\, n(r)/r$
should be considered. However, these corrections are
$\mathcal{O}(\delta/R)$ relative to the term $p_1+m_p\, n_1\,
\tilde{v}_1^2$ when $\delta/R\ll 1$ applies; in particular, we
have checked this to hold for any reasonable DM and gas
distributions $M(<r)$ and $n(r)$ inserted in the above integral.

In fact, for the numerous small merging partners with mass ratio
$M'/M\la 5\%$ the diffuse baryonic component is stripped away
promptly (i.e., within a layer $\delta/R\la 10\%$) from its DM
counterpart, and raised to the final temperature and entropy
levels (see Tormen et al. 2004). These smaller merging partners
with cooler $T'_v$ undergo prompter and also stronger transitions;
so they not only are better described by Eqs.~(B2), (B3) and (B4),
but also dominate the averaging procedure over the distribution of
progenitor masses $M'$, that are used in the main text to compute
the effective value of $K_2(T)$. For all these reasons, we refer
to ``shocks'', but imply that similar results also hold for thin
layers.

It is useful to relate the inflow velocity $\tilde{v}_1$ in the
\emph{shock} rest frame to the infall velocity $v_1$ in the
\emph{cluster} rest frame, to obtain
\begin{equation}
\tilde{v}_1={2\over 3}\, v_1\, \left[1+\sqrt{1+{15\over 4}\, {k
T_1\over \mu m_p\, v_1^2}}\, \right]~;
\end{equation}
here we have assumed the kinetic energy to be small downstream. In
turn, the infall velocity $v_1$ is set by the DM gravitational
potential $\Phi_2$ at the virial radius through energy conservation,
which yields
\begin{equation}
v_1 = \sqrt{-2\, \chi\, \Phi_2}~;
\end{equation}
the fudge parameter $\chi = 1 - R/R_f$ expresses the ignorance on
the exact position of the radius $R_f\ga R$ at which free fall
begins (see Voit et al. 2003). The upper bound $\chi\approx 0.7$
obtains if $R_f$ is computed by equating the free-fall speed to
the Hubble flow. In fact, basing on many numerical simulations
since Bertschinger (1985) we adopt the value $\chi\approx 0.37$,
close to that obtained if the gas inflow begins at the turnaround
radius during the gravitational collapse of a standard ``top-hat''
perturbation (see Padmanabhan 2003).

We now derive a number of convenient approximations to the
temperature and density jumps valid for very strong, moderately
strong, and weak shocks, that are used in \S~4 of the main text. For
very strong shocks with $kT_1/\mu m_p\, v_1^2\ll 1$ that occur for
supersonic accretion onto rich clusters, Eq.~(B5) reads
$\tilde{v}_1\simeq 4\, v_1/3$; thus Eq.~(B2) and Eq.~(B4)
respectively approximate to
\begin{equation}
\theta\simeq {1\over 3}\,{\mu m_p v_1^2\over kT_1} ~~~~~~~~ {n_2\over n_1}\simeq 4 ~;
\end{equation}
the former expresses the limit where full thermalization of the
inflow kinetic energy takes place. We can now use Eq.~(B6) with
$\Phi_2 \approx -5.7\, \sigma^2$ (the NFW potential corresponding to
a concentration $c=5$) to find $v_1\approx 2.1\, (kT_v/\mu
m_p)^{1/2}$; putting all together yields $\beta= kT_v/\theta\,
kT_1\simeq 3\, kT_v/\mu m_p\, v_1^2\approx 0.7$, a value consistent
with the observations of rich clusters.

For moderately strong shocks with $kT_1/\mu m_p\, v_1^2\la 1$ that
occur for preheated accretion onto poor clusters, Eq.~(B5) gives
$\tilde{v}_1\simeq 4\, v_1/3 + 5\, kT_1/4\, \mu m_p\, v_1$. Now
Eqs.~(B2) and (B4) approximate to
\begin{equation}
\theta\simeq {1\over 3}\,{\mu m_p v_1^2\over kT_1} + {3\over 2} ~~~~~~~~
{n_2\over n_1}\simeq 4\, \left(1-{15\over 16}\, {1\over \theta}\right) ~;
\end{equation}
relatedly, the parameter $\beta=T_v/\theta\, T_1$ is lowered from
the cluster value around $0.7$ to
\begin{equation}
\beta\simeq 0.7\, \left(1-{3\over 2}\, {1\over \theta}\right)~.
\end{equation}

For weak shocks with $kT_1/\mu m_p\, v_1^2\gg 1$ that occur for
preheated accretion in small groups, one recovers from Eqs.~ (B2)
and (B4) the adiabatic limit
\begin{equation}
\theta\simeq 1 + \sqrt{4\, \mu m_p\, v_1^2\over 15\, kT_1}~~~~~~~~ {n_2\over n_1}\simeq 1+{3\over 2}\, (\theta-1)~.
\end{equation}

\section{A new family of self-similar blast waves}

Hydrodynamic flows are amenable to a self-similar description when
their dynamics can be characterized in terms of space and time
variables, and of a small set of parameters with independent
dimensions. Although self-similar solutions are only particular
descriptions of a hydrodynamic flow, they often accurately yield
its actual asymptotic behavior and offer a useful guide for
understanding its generic features, as discussed by Zel'dovich \&
Raizer (1967).

Self-similar solutions have been systematically investigated by
Sedov (1959) for blast waves originated by time-dependent energy
discharges into a gas with power-law density. Such solutions have
then been successfully applied to a variety of astrophysical
problems, such as the propagation of SN remnants (see Chevalier
1976), and of shocks driven by solar flares (see Parker 1963).
Most of these treatments consider a gas with negligible initial
pressure so that the resulting blast wave is strongly supersonic;
moreover, they do not include gravitational effects, in particular
those due to a dominant DM component.

In the main text we are interested in \emph{strong} and \emph{weak}
blasts driven by comparable energy discharges into plasmas in virial
equilibrium with the DM at temperatures differing by factors $10^2$
from galaxies to clusters. To cover the full range, we derive here
(see also Lapi 2004) a new family of self-similar solutions that not
only include a power-law initial density gradient and time-dependent
energy injection, but also incorporate DM gravity and a finite
initial pressure. In Fig.~5 we preliminarily outline the unperturbed
equilibrium and the perturbed flow.

\subsection{The ambient medium}

We consider initial configurations constituted by DM profiles
\begin{equation}
\rho(r)=\rho_R\, \left({R\over r}\right)^{\omega}
\end{equation}
with power-law index $2\leq\omega < 5/2$; the case $\omega=2$
corresponds to the standard isothermal sphere. The plasma is
assumed to be in equilibrium within the DM potential well with
initial density $n (r)\propto \rho(r)$. Such simple power-law runs
will be useful for obtaining self-similar blasts, but also
constitute fair piecewise approximations of the plasma
distributions expressed by Eq.~(2). The plasma temperature and
entropy profiles are given by
\begin{eqnarray}
&\nonumber & \frac{kT}{\mu m_p} = {2\pi G \, \rho_R
R^{\omega}\over (\omega-1)\, (3-\omega)}\, r^{2-\omega}~,\\
& \nonumber &\\
&\nonumber & K = {2\pi G\, \mu m_p^{5/3}\, (\rho_R
R^{\omega})^{1/3}\over (\omega-1)\, (3-\omega)\, (m/M)^{2/3}}\, r^{2-\omega/3
}~;
\\
\end{eqnarray}
relatedly, the polytropic index $\Gamma = 2\, (1-1/\omega)$ ranges
from $1$ to $1.2$ when $\omega$ increases from $2$ to $2.5$. The
total energy of the gas in equilibrium is
\begin{equation}
E_{\mathrm{tot}}(<r) = {1\over 2}\, \left[-{3\over 5}\,
{3-\omega\over 5-2\, \omega}\, (4\, \omega-7) \right]\, m(<r)\, c_s^2\propto r^{5-2\, \omega}~;\\
\end{equation}
here $m(<r) = 4\pi\, m_p\, \int^r{dx}\, x^2\, n(x) = 4\pi \, (m/M)\,
\rho_R R^{\omega}\, r^{3-\omega}/(3-\omega)$ is the gas mass within
$r$, and the sound speed is $c_s = (5\, k T/3\, \mu
m_p)^{1/2}\propto r^{1-\omega/2}$. The bounds $2\leq\omega<2.5$
guarantee the total energy of the system to be finite and negative;
hereafter and in the main text we indicate its modulus with
$E(<r)\equiv -E_{\mathrm{tot}}(<r)$.

\subsection{The blast}

A blast wave sweeps through the plasma as a result of the energy
injections $\Delta E(t)$ by a central source. The ensuing unsteady
gas flow is described by the system of partial differential
equations
\begin{eqnarray}
&\nonumber &\partial_t{n}+\partial_r{(n v)}+{2\, n v\over r} = 0\\
&\nonumber &\\
& \nonumber &\partial_t{v}+v\partial_r{v}+{1\over m_p\, n}\,\partial_r{p}+{G\, M(<r)\over
r^2}=0\\
&\nonumber &\\
&\nonumber & (\partial_t{}+ v \partial_r{})\, {p \over n^{5/3}}
= 0~,\\
\end{eqnarray}
supplemented at the leading shock by the Rankine-Hugoniot boundary
conditions. The latter may be derived from the general expressions
Eqs.~(B1) specialized to the case of \emph{internal} shocks, i.e.,
taking the preshock flow velocity $\tilde{v}_1$ in the shock rest
frame equal to $- v_s$, opposite to the shock velocity. This yields
\begin{equation}
v_2 = {3\over 4}\, v_s\, {\mathcal{M}^2-1\over \mathcal{M}^2}~~~~~
{p_2\over p_1} = {5\, \mathcal{M}^2-1\over 4}~~~~~ {n_2\over
n_1} = {4\, \mathcal{M}^2\over \mathcal{M}^2+3}~,
\end{equation}
in terms of the Mach number $\mathcal{M}\equiv v_s/c_s$. Note that
the condition for shock formation and propagation $v_s\geq c_s$
can also be written as $p_2+m_p\, n_2\, v_2^2\geq p_1$, i.e., the
total stress pushing the shock outward has to exceed the upstream
pressure.

The flow described by the above equations will be self-similar if it
can be expressed in terms of the variables $r$ and $t$ and of only
two more parameters with independent dimensions. One must be the
gravitational constant $G$ if gravity effects are to be included; as
to the other, we take the quantity $\rho_R R^{\omega}$ entering the
initial state given by Eq.~(C1).

Self-similarity then implies for the law of energy injection
\begin{equation}
\Delta E(t) \propto G^{(5-\omega)/\omega}\, (\rho_R
R^{\omega})^{5/\omega}\, t^{2\, (5-2\, \omega)/\omega}~.
\end{equation}
The resulting time-dependencies of the source output turn out to
be interesting; the values of the index $\omega$ correspond to
luminosities going from a \emph{constant} ($\omega=2$) to a
\emph{spike} ($\omega=5/2$), the upper range being useful to
describe the quasar fading out because of its own feedback on the
accreting gas. Since $E(<R_s)\propto R_s^{5-2\, \omega}\propto
t^{2\, (5-2\, \omega)/\omega}\propto \Delta E$ holds, it is easily
seen that $\Delta E/E$ is constant during the blast motion, and
thus constitutes the key parameter for labeling the solutions.
This comes about because the values of $\Delta E/E$ sets the Mach
number, i.e., the strength of the shock, as we specify below.

Under self-similarity, Eqs.~(C4) are solved along the following
lines. First, we use the dimensional parameters of the problem to
construct the self-similarity (adimensional) variable $\xi \equiv
r/R_s(t)$, where
\begin{equation}
R_s (t) = \left[{5\pi G\, \rho_R R^{\omega}\, \omega^2 \,
\mathcal{M}^{2} \over 6\, (\omega-1)\, (3-\omega)}\right]^{1/
\omega}\, t^{2/\omega}\propto R\, (\mathcal{M}\, t)^{2/\omega}
\end{equation}
is the shock radius. It is now seen that the Mach number
$\mathcal{M}=v_s(t)/c_s[R_s(t)]$ is independent of time and
position, as is $\Delta E/E$. Note that in our simple model with
$\omega=2$ the blast moves out with constant speed, while for
$\omega>2$ it decelerates.

\begin{figure}
\epsscale{.6}\plotone{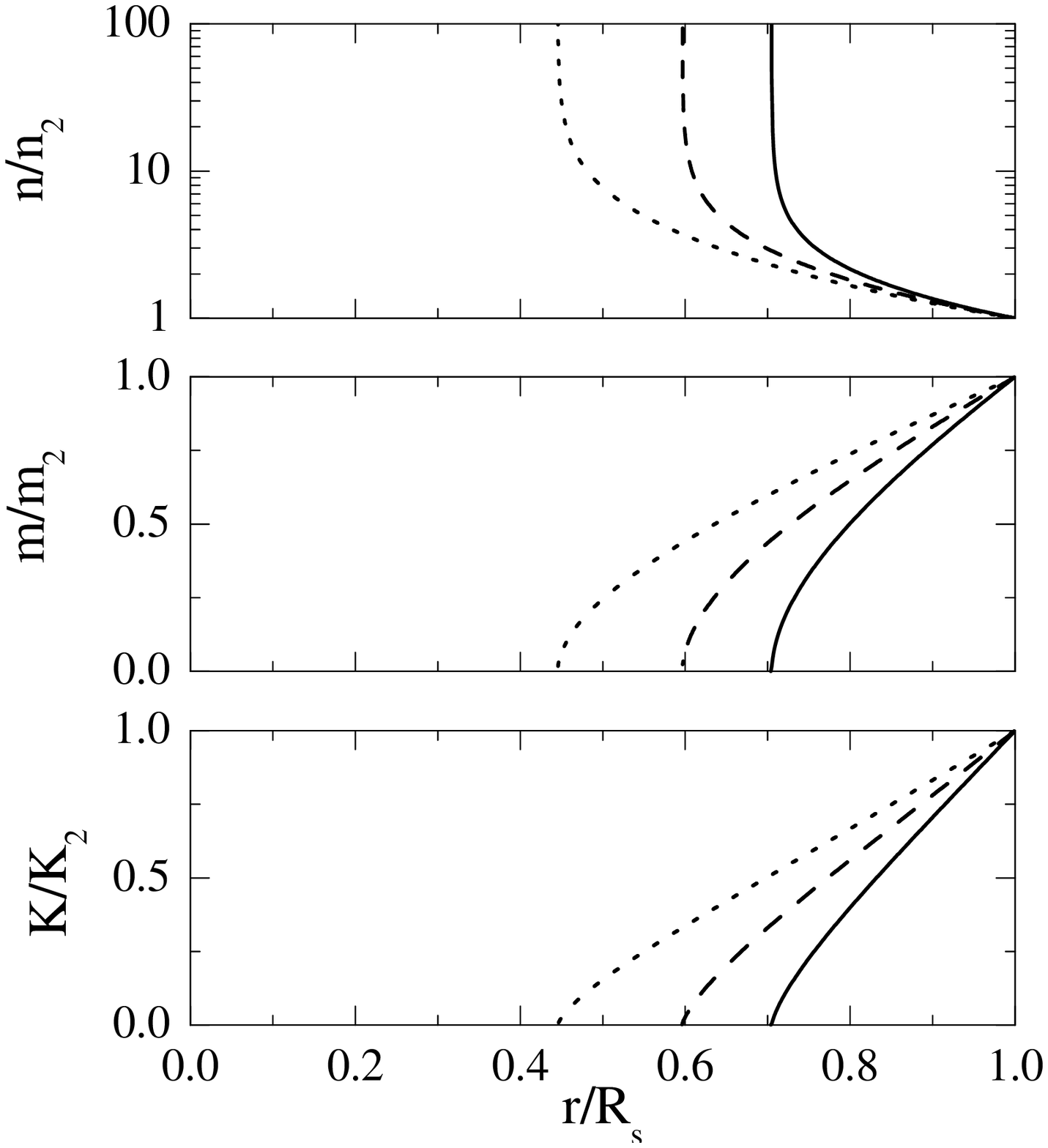}\caption{Distributions of density
(\textit{top panel}), cumulative mass (\textit{middle panel}) and
entropy (\textit{bottom panel}) within the blast, computed from
Eqs.~(C9) and (C10) for $\omega=2$, and normalized to their
postshock values. \textit{Solid} lines are for a strong shock with
$\Delta E/E=3$, \textit{dashed} lines for an intermediate shock with
$\Delta E/E=1$, and \textit{dotted} lines for a weak shock with
$\Delta E/E=0.3$.}
\end{figure}

Then we introduce the adimensional quantities $\mathcal{V}(\xi)$,
$\mathcal{D}(\xi)$, and $\mathcal{T}(\xi)$ through
\begin{equation}
v(r,t)={r\over t}\, \mathcal{V}(\xi)~~~~~~~~ n(r,t)={1\over m_p\, G\,
t^2}\, \mathcal{D}(\xi)~~~~~~~~ T(r,t)={3\, \mu m_p\over 5\,
k}{r^2\over t^2}\, \mathcal{T}(\xi)~;
\end{equation}
these enable us to convert the partial differential Eqs.~(C4) into a
set of ordinary differential equations
\begin{eqnarray}
& \nonumber & \xi\, \left[\mathcal{V}'+\left(\mathcal{V}-{2\over \omega}\right)\, {\mathcal{D}'\over \mathcal{D}}\right]=2-3\, \mathcal{V}\\
& \nonumber &\\
&\nonumber &\xi\left[\mathcal{V}'\, \left({2\over
\omega}-\mathcal{V}\right)-{3\over 5}\, \mathcal{T}\,
\left({\mathcal{T}'\over \mathcal{T}}+{\mathcal{D}'\over
\mathcal{D}}\right)\right]={6\over 5}\,
\mathcal{T}+\mathcal{V}^2-\mathcal{V}+{24\over 5}\, {\omega-1\over
\omega^2\mathcal{M}^2}\,
\xi^{-\omega}\\
& \nonumber &\\
&\nonumber & \xi\, \left(\mathcal{V}-{2\over \omega}\right)\,
\left({\mathcal{T}'\over \mathcal{T}}-{2\over 3}\,
{\mathcal{D}'\over \mathcal{D}}\right)=-2\,
\left(\mathcal{V}-{1\over
3}\right)~,\\
\end{eqnarray}
with boundary conditions at $\xi=1$ ($r=R_s$) given by
\begin{equation}
\mathcal{V}_s= {3\over 2}\,
{\mathcal{M}^2-1\over
\omega\, \mathcal{M}^2}~;~~~\mathcal{T}_s = {(5\,
\mathcal{M}^2 - 1)\, (\mathcal{M}^2+3)\over 4\, \omega^2\, \mathcal{M}^4}~;~~~
\mathcal{D}_s= {24\, (m/M)\over 5\,
\pi}\, { (3-\omega)\, (\omega-1)\over \omega^2\,
(\mathcal{M}^2+3)}~.
\end{equation}

As a last step, we have numerically solved the differential system
Eqs.~(C9) by using a standard Runge-Kutta integrator with
adjustable time step. For various shock strengths we show in
Fig.~C1 the distributions of density, mass, and entropy in the
blast for the $\omega=2$ model.

While the radial mass and entropy distributions differ considerably
in the blast and in the initial configuration, the entropy
distribution $\kappa(m)$ as a function of the integrated plasma mass
$m(<r)$ remains unchanged, and reads
\begin{equation}
\kappa(m)\propto m(<r)^{(6-\omega)/3\, (3-\omega)}~.
\end{equation}
The postshock normalization $K_2$ is raised because of dissipation
within the shock, after the \emph{general} Eq.~(5) in the main
text that has been derived in Appendix B.

\begin{figure}
\epsscale{.6}\plotone{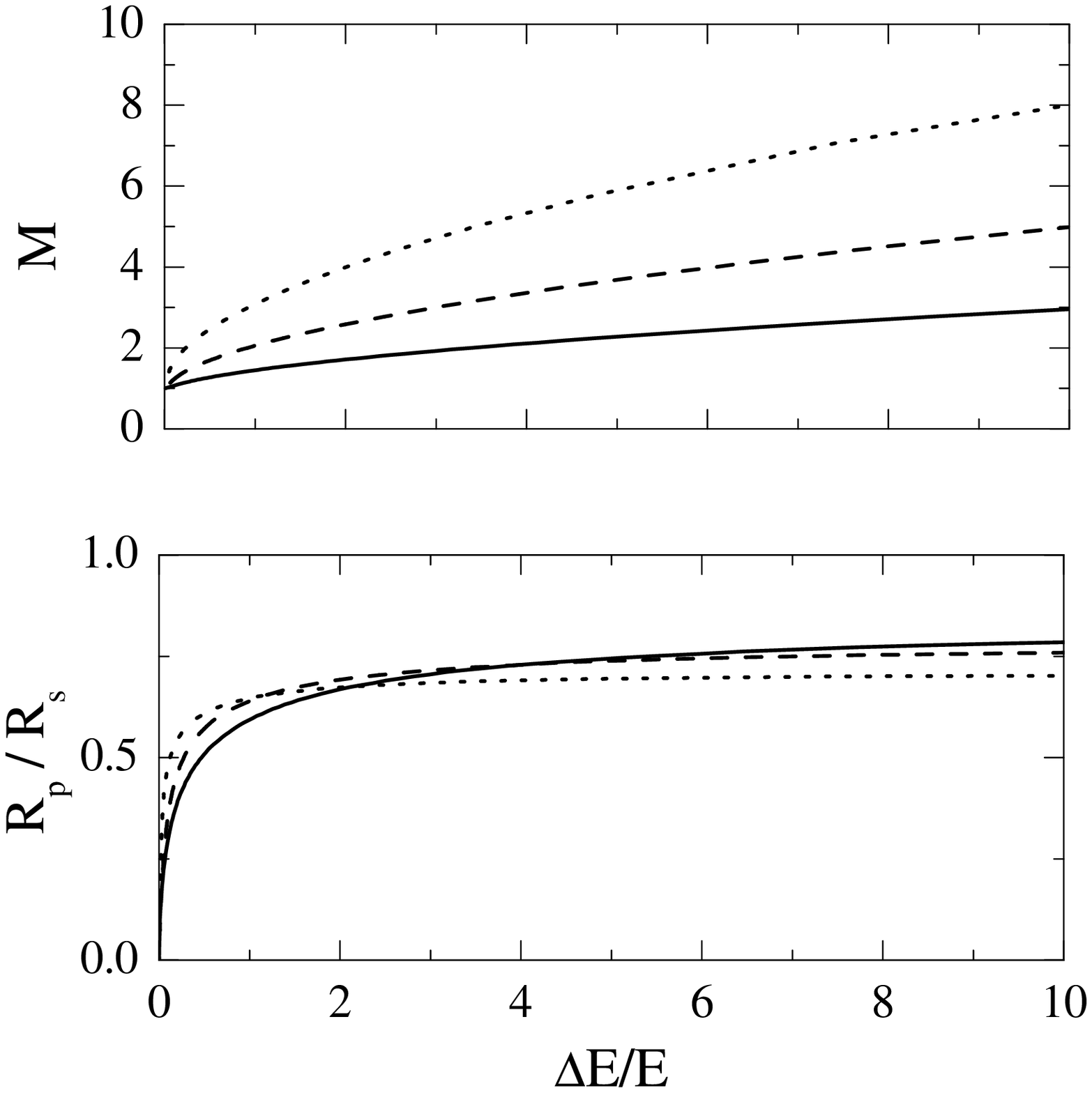}\caption{The Mach number $\mathcal{M}$
(\textit{top panel}) and the piston position $\lambda\equiv R_p/R_s$
(\textit{bottom panel}) as a function of $\Delta E/E$. The
\textit{solid} line is for $\omega=2$, the \textit{dashed} line is
for $\omega=2.25$, and the \textit{dotted} line is for
$\omega=2.4$.}
\end{figure}

The overall energy balance for the blast flow is written
\begin{equation}
\Delta E(t) - E(<R_s) = 4\pi\,
\int_{0}^{R_s}{dr}\, r^2\,
\left\{{1\over 2}\, m_p\, n\, v^2 + {3\over 2}\, p -
{G\, M(<r)\over r}\, m_p\, n\right\}~.
\end{equation}
This equation relates $\mathcal{M}$ and $\Delta E/E$, and the
result is shown in the top panel of Fig.~C2. Note that for strong
shocks $\Delta E/E\propto \mathcal{M}^2$ holds; in particular,
such a dependence implies that in the limit $\mathcal{M}^2\gg 1$
(but in fact already for $\mathcal{M}^2>3$) our family of
self-similar blast waves converge to the standard solutions with
negligible gravity and zero initial gas pressure. E.g., for
$\omega = 2$ and constant source luminosity one has $R_s \propto
\mathcal{M}\, t\propto (\Delta E/E)^{1/2}\, t$; since $E\propto
R_s$ holds, one recovers the shock motion $R_s\propto L^{1/3} t
\propto \Delta E^{1/3}\, t^{2/3}$, provided by the standard blast
wave theory in the strong shock limit. As another example,
consider $\omega=2.5$ and spiky energy liberation after Eq.~(C6),
for which one has $R_s \propto \mathcal{M}^{4/5}\, t^{4/5}\propto
(\Delta E/E)^{2/5}\, t^{4/5}$; since now $E=$ const holds, one
recovers the standard dependence $R_s\propto \Delta E^{2/5}
t^{4/5}$ for strong shock.

From Fig.~C1 it is easily seen that the flow is confined within a
shell that terminates at the leading shock at $R_s$ and begins at
a trailing ``piston'', the contact discontinuity located at $R_p =
\lambda\, R_s < R_s$ where the action of the source is transferred
to the plasma. Self-similarity implies the thickness $1-\lambda$
of such a shell to depend only (and inversely) on the shock
strength; $\lambda$ is plotted vs. $\Delta E/E$ in the bottom
panel of Fig.~C2.

Analytic approximations may be derived for the limiting behavior of
the adimensional variables $\mathcal{V}$, $\mathcal{D}$,
$\mathcal{T}$ in the vicinity of the piston. For a given $\omega$
such limiting behaviors turn out to be independent of the shock
strength, and read
\begin{equation}
\mathcal{V}_p \simeq {2\over \omega}-{6\over 5}\, \left({7\over
\omega}-2\right)\, \left[{\xi\over \lambda}-1\right]~;~~~~
\mathcal{D}_p\propto \left[{\xi\over
\lambda}-1\right]^{(\omega-6)/3(7-2\, \omega)}~;~~~~ \mathcal{T}_p
\propto {1\over \mathcal{D}_p}~.
\end{equation}
Thus at the inner piston the density diverges weakly but the mass
vanishes (so the overall effects of radiative cooling are
negligible), while the temperature goes to zero making up a finite
pressure.

\subsection{The shell approximation}

Since the perturbed flow is confined within a shell of constant
thickness $\lambda$ between the inner piston and the leading
shock, it is convenient to represent our solutions by using the
\emph{shell} approximation (see Cavaliere \& Messina 1976;
Ostriker \& McKee 1988). Here we improve on the classic treatment
by extracting the value of the shell thickness $\lambda$ directly
from the exact solution (see Fig.~C2), in order to obtain results
reliable to better than $15\%$.

The equation of motion for the shell is written
\begin{equation}
{d\over dt}\left[m(<R_s)\, v_2\right] = 4\pi\, R_s^2\, \left[\langle p \rangle\,
-p_1\right]-{3-\omega\over 5-2\, \omega}\, {G\, M(<R_s) \over R_s^2}\, m(<R_s)~;
\end{equation}
here $\langle p \rangle$ is the volume-averaged pressure, which as
function of $\mathcal{M}$ reads
\begin{equation}
{\langle p \rangle\over p_1} = {5\over 8}\, {8-3\, \omega\over
3-\omega}\, (\mathcal{M}^2-1)+ {3\over 5-2\, \omega}~.\\
\end{equation}
For weak shock with $\mathcal{M}\rightarrow 1$ we obtain $\langle
p \rangle/p_1 = 3/(5-2\, \omega)$, while in the strong shock limit
$\mathcal{M}\gg 1$ our result $\langle p \rangle/p_1 \rightarrow
5\, \mathcal{M}^2\,(8-3\, \omega)/8\, (3-\omega)$ matches that
known for standard, strong blast waves without gravity (e.g.,
Cavaliere \& Messina 1976).

Integrating Eq.~(C14) leads to
\begin{equation}
\Delta E(t) - E(<R_s) = {1\over 2}\, m(<R_s)\, v_2^2 + {3\over 2}\, \langle
p\rangle\, V-{3-\omega\over 5-2\, \omega}\, {G\, M(<R_s)\over R_s}\, m(<R_s)~,
\end{equation}
in terms of the shell volume $V = 4\pi R_s^3(1-\lambda^3)/3$; this
is the simplified, shell version of Eq.~(C12).

A relevant quantity is the ratio between the kinetic and the
thermal energy of the blast; inserting Eq.~(C15) into Eq.~(C16)
one finds
\begin{equation}
{\Delta E_{\mathrm{Kin}}\over \Delta E_{\mathrm{Th}}} = {3\over
2\, (8-3\,\omega)\, (1-\lambda^3)}\, {\mathcal{M}^2-1\over
\mathcal{M}^2}~.
\end{equation}
This is easily seen to vanish in the weak shock limit
($\mathcal{M}\rightarrow 1$), and to take on values $3/(16-6\,
\omega)\, (1-\lambda^3)$ for strong shocks with $\mathcal{M}\gg
1$, as in standard blasts with no gravity.

Finally, we remark that Eqs.~(C7), (C15), (C16), and (C17)
constitute a set of handy relations useful in the main text and
suitable to implement in semianalytic models (SAMs) or
hydro$+N$-body codes of galaxy formation.

\end{appendix}

\end{document}